\documentclass[fleqn,usenatbib]{mnras}

\usepackage{mathptmx}
\usepackage{txfonts}

\usepackage[T1]{fontenc}
\usepackage{ae,aecompl}

\usepackage{lastpage}
\usepackage{graphicx}	
\usepackage{amssymb}	
\usepackage[usenames, dvipsnames]{color}
\usepackage{comment}
\usepackage{cancel}

\title[Sub-arcsec source counts at low frequencies]{Interplanetary Scintillation studies with the Murchison Wide-field Array III: Comparison of source counts and densities for radio sources and their sub-arcsecond components at 162\,MHz}
\author[R. Chhetri et al.] {R. ~Chhetri$^{1,2}$\thanks{email:rzn.chhetri@gmail.com}, R. D. ~Ekers $^{1, 3}$, J. Morgan$^{1,2}$, J.-P. Macquart$^{1,2}$, T. M. O. Franzen$^{4,5}$ \\
$^{1}$International Centre for Radio Astronomy Research, Curtin University, GPO Box U1987, Perth, WA 6845, Australia\\
$^{2}$ARC Centre of Excellence for All-Sky Astrophysics (CAASTRO), Australia\\
$^{3}$CSIRO Astronomy and Space Science (CASS), Marsfield, NSW 2122, Australia\\
$^{4}$CSIRO Astronomy and Space Science, PO Box 1130, Bentley WA 6102, Australia\\
$^{5}$Netherlands Institute for Radio Astronomy (ASTRON), PO Box 2, 7990 AA Dwingeloo, The Netherlands
}
\date{Accepted XXX. Received YYY; in original form ZZZ}

\pubyear{2018}

\begin{document}
\label{firstpage}
\pagerange{\pageref{firstpage}--\pageref{lastpage}}
\maketitle

\begin{abstract}
We use Murchison Widefield Array observations of interplanetary scintillation (IPS) to determine the source counts of point ($<$0.3\,arcsecond extent) sources and of all sources with some subarcsecond structure, at 162\,MHz. We have developed the methodology to derive these counts directly from the IPS observables, while taking into account changes in sensitivity across the survey area.
The counts of sources with compact structure follow the behaviour of the dominant source population above $\sim$3\,Jy but below this they show Euclidean behaviour. We compare our counts to those predicted by simulations and find a good agreement for our counts of sources with compact structure, but significant disagreement for point source counts. Using low radio frequency SEDs from the GLEAM survey, we classify point sources as Compact Steep-Spectrum (CSS), flat spectrum, or peaked. If we consider the CSS sources to be the more evolved counterparts of the peaked sources, the two categories combined comprise approximately 80\% of the point source population.
We calculate densities of potential calibrators brighter than 0.4\,Jy at low frequencies and find 0.2 sources per square degrees for point sources, rising to 0.7 sources per square degree if sources with more complex arcsecond structure are included. We extrapolate to estimate 4.6 sources per square degrees at 0.04\,Jy. We find that a peaked spectrum is an excellent predictor for compactness at low frequencies, increasing the number of good calibrators by a factor of three compared to the usual flat spectrum criterion.

\end{abstract}




\section{Introduction}

There has been a resurgence of interest in low frequency radio surveys spawned by recent, sensitive wide-field facilities offering the opportunity to shed new light on the evolutionary history of the extragalactic radio source population.  One crucial aspect in unravelling the evolution of the various sub-populations is angular resolution. The ability to resolve structure on arcsecond scales affords the possibility to distinguish between highly extended radio galaxies, those whose megaparsec-scale emission is the hallmark of ancient nuclear activity, and the highly compact population, whose emission confined to the inner few kiloparsecs of the galaxy signifies more recent nuclear activity.  These populations are expected to exhibit differing evolutionary characteristics, as indicated by statistical censuses of the flat- and steep-spectrum populations observed at gigahertz frequencies \citep[e.g.][]{Condon1984a}.

The number count statistics furnished by low frequency radio surveys also address more prosaic but fundamental questions related to the calibratability of forthcoming radio facilities, particularly SKA\_LOW which requires a high density of compact sources to provide the direction dependent calibration. Here, again, high angular resolution information is pivotal in determining the areal density of sources compact enough to serve as effective phase calibrators.   

The counts of the overall source population at low radio frequencies have been measured to high precision over a wide range of flux density. For example, \cite{Intemaetal2011} and \cite{Williams_2013A&A...549A..55W} derived the radio source count statistics from separate deep observations of the Bootes field, reaching down to $\sim$10 mJy at 153\,MHz. Recently, \cite{Intema2017A&A...598A..78I} derived more precise counts from wide area using the TGSS ADR1 data down to 0.1\,Jy at 150\,MHz. At higher flux densities Franzen et al. (2018, submitted) have produced extragalactic source counts from a very wide (24831\,sq.\,deg.) region of the sky at 88, 118, 154 and 200\,MHz from the Galactic and Extragalactic MWA (GLEAM) catalogue \citep{Hurley-Walker2017}. Since the radio population is dominated by extended radio sources at low radio frequencies, their corresponding source counts primarily examine the characteristics of the extended radio source population. 

To date there has been a dearth of information on the compact source population at low frequencies. \cite{Readhead_1975MNRAS.170..393R} used interplanetary scintillation (IPS) to identify those objects with sub-arcsecond compact emission at 81.5\,MHz, and examine their counts at flux densities $>$2\,Jy.  Their catalogue was based on a sample drawn from the 3C survey, which is complete to 9\,Jy at 178\,MHz.  Their statistics revealed that the strongly scintillating (i.e. highly compact, arcsecond-scale) sources exhibit strong cosmological evolution of the form inferred for quasars and powerful radio sources in general. \cite{Artyukh_1996ARep...40..601A} made IPS observations at 102 MHz of a sample derived from the 7C survey at 151 MHz.  They reported the flux density distribution of the compact components of their sources down to a flux density threshold of 0.1\,Jy, and claimed a sharp drop in the prevalence of scintillating sources at low flux densities. However, this claim is problematic because the dropoff was at flux densities well below their reported completeness limit.

More recently, the Murchison Widefield Array has revitalised  the study of the extragalactic sub-arcsecond compact source population at low radio frequencies in significant numbers using IPS \citep{Morgan_2018MNRAS.473.2965M}.  By imaging each $\sim 900\,$sq.\,deg field with 0.5\,s cadence, it is able to effectively identify, via IPS, the sub-arcsecond source population, and undertake large-sample statistical investigations into its properties. This technique has now demonstrated its utility as a potent probe of the low-frequency radio population \citep{Chhetri_2018MNRAS.474.4937C}. Optical identifications of sub-arcsecond compact objects in \cite{Chhetri_2018MNRAS.474.4937C} are discussed in Sadler et al. (2018, submitted).  

In this paper, we examine the differential source count distribution of the sub-arcsecond radio population to $\sim$0.4 Jy at 162 MHz using the data presented in \cite{Chhetri_2018MNRAS.474.4937C}. The source identification and IPS studies were both conducted at the same observing frequency and epoch, obviating the ambiguities of previous studies imposed by the necessity of interpreting source catalogue information at a frequency substantially different from the frequency at which the IPS measurements were conducted. In Section \ref{Sec:Method} we present the details of the parent sample used in our study and our methodology, and derive the source count statistics. In Section \ref{Sec:Discussion} we discuss the implications of our results, and our conclusions are presented in Section \ref{Sec:Conclusions}. Throughout this work we follow the convention of $S_{\nu} \propto \nu^{\alpha}$, where $S_\nu$, $\nu$ and $\alpha$ are flux density, frequency and spectral index respectively.

\section{Method}
\label{Sec:Method}

\subsection{Data}
We derive the source counts from the $\sim 900\,$sq.deg. MWA field presented in \cite{Chhetri_2018MNRAS.474.4937C}, made using a single 5-minute observation centred at 162\,MHz. This field contains 2550 objects at a signal-to-noise (S/N) ratio $\geq$5, which we shall refer to as the ``Total Population". 

We determined the magnitude of the interplanetary scintillation (IPS) exhibited by our sources by measuring the standard deviation of all pixels in images produced at a half-second cadence to produce a ``variability image''.   Further details of the variability image are discussed in \citet[][]{Morgan_2018MNRAS.473.2965M} and \citet{Chhetri_2018MNRAS.474.4937C} (hereafter, Papers I and II respectively).  Following the convention established in Paper I, we term the continuum image (i.e. the mean flux density image) the ``standard image".

In \cite{Chhetri_2018MNRAS.474.4937C}, we derived the normalised scintillation indices (NSIs) for all sources in the field. The NSI gives the fraction of flux density that is concentrated on sub-arcsec scales. The relevant angular scale probed by IPS, the Fresnel scale, is $\sim$ 0.3 arcsecond for our field at 162\,MHz. By imposing the criterion NSI$\geq$0.90 we select compact sources whose flux is almost entirely confined to $<$ 0.3 arcsecond angular scales. Of the 247 ``Sources with compact structure" that exhibit IPS with a S/N$\geq 5$ in the variability image, there are 93 sources (the ``Point source population'') that meet this NSI criterion.  Note that in contrast to Paper II we only consider those sources for which there is a detection of a point source (aside from the validation of our method in Section~\ref{Sec:noCountTotalPop}). The upper limits on NSI used in Paper II are only relevant when comparisons are being drawn between compact and non-compact sources, which is not relevant to the the work presented here. This also allows us to use the full sample of compact sources, rather than restricting ourselves to a high S/N subsample.

\subsection{Source counts}
There are three inherent complications to the analysis of the source counts in this dataset: (i) there is a large gradation in sensitivity across the field due to the attenuation of the primary beam; (ii) the scintillation index (and therefore point source sensitivity) are also a function of solar elongation and; (iii) derivation of the compact source counts relies on the interpretation of the statistics of the standard deviation of the flux density, rather than the mean flux density. Point (i) is discussed directly below, while discussion of points (ii) and (iii), being IPS-specific, are deferred to Section \ref{Sec:noCountCompactSrc}.

\subsubsection{Total population source counts}
\label{Sec:noCountTotalPop}

An important consideration when deriving the source count statistics from a flux-limited survey is the estimation of area of the sky probed at a given flux density limit. This is straightforward if the survey area has constant sensitivity.  However, images made with single pointing of the MWA vary greatly in sensitivity across the wide field of view, according to the primary beam. 

Estimation of the source counts in the present case is performed by considering the survey to be the ensemble of a number of surveys with differing completeness limits. The source counts are then visualised as the combination of source counts drawn from surveys with different flux density limits. The approach is to determine the area of the image sensitive to a certain flux density limit, and then derive the source counts using the area that corresponds to this particular limit. 

In order to estimate the area sensitive to a given flux density, we produced a ``sensitivity image".  This requires two inputs (i) an image of the noise (i.e.\,\,the ``RMS image'') derived from our standard image, and (ii) the primary beam model.  The RMS image is output by the source-finding tool Aegean \citep{Hancock2012} and contains the values for root-mean-square (RMS) in the Gaussian background noise estimated for each pixel in the input image. 

The primary beam correction to our image\footnote{The primary beam correction was not necessary for the purposes of the analysis performed in Paper II.}, was made using an appropriately scaled and weighted primary beam model, as described in \cite{Hurley-Walker_2014PASA...31...45H} and \cite{Sault_1996A&AS..120..375S}.  We note that the linearly polarised primary beams of the MWA differ in sensitivity across the sky.  We used primary beam model of \cite{Sutinjo_2015} and \cite{Sokolowski_2017PASA...34...62S}, and then made primary beam corrections to each pixel in the noise image and applied an absolute flux scale correction to each pixel. 
We then multiplied each pixel in the primary beam corrected noise image by the detection limit for our standard image, namely S/N=5. 

We thus obtained a nominal noise floor image, the sensitivity image, from which it is possible to estimate the area over which we are sensitive to sources at any given flux density level. For ease of estimation, we performed all the above steps on the RMS image with zenith equal area projection applied, such that each pixel in the image subtends an equal area ($28\arcsec.8\, \times \, 28\arcsec.8$) on the sky. 

The sensitivity image simplifies the estimation of the survey area for a given flux density. For instance, to estimate the total survey area over which a source with flux density of 1 Jy could be detected at S/N $\geq$ 5, we simply count the number of pixels with values below 1\,Jy in the sensitivity image. The resulting number of pixels multiplied by the area subtended by each pixel is the survey area for sources at that flux density threshold.

Standard source count analysis involves determining the number of sources per unit area in each of a series of logarithmic flux density bins.
In order to account for the fact that the survey area changes \emph{across} each of our bins, we first calculated $\Omega_i$: the area (in steradians) over which each source $i$ could have been detected.
The number of sources per steradian $dN$ for each bin was then calculated as $\sum_{i=0}^{N} \Omega_i^{-1}$, where the summation is over all of the sources in the bin.

To convert the source counts into the Euclidean weighted differential form (i.e. $S^{2.5}\frac{dN}{dS}$), we took $S$ to be the median flux density of all sources in each bin. $dS$ is simply the bin width in Jy.

\subsubsection{Validation of the source counts method against GLEAM statistics}

Franzen et al. (2018, submitted) have derived the source counts for the total GLEAM survey populations at 88, 118, 154 and 200 MHz. Their counts at 154\,MHz reaches down to 80\,mJy. To make direct comparisons against our counts performed at 162\,MHz, we extrapolated their 154\,MHz counts to 162\,MHz using their result that, for $S_{154\,{\rm MHz}}> 0.5\,$Jy, a spectral scaling index of -0.8 provides an excellent conversion (four percent) of the source counts between these slightly different frequencies. 

We followed the prescription of \citet[][equation 8]{Franzen2015MNRAS.453.4020F} to identify objects as either extended or point-like at the resolution of the MWA synthesized beam at 162 MHz, such that the appropriate flux density, either integrated or peak flux density, is used. This ensured that our source counts are estimated in a manner identical to the GLEAM counts, enabling a direct comparison between the two.  Based on this approach, we used the integrated flux density for 239 out of 2550 (9.3\%) objects in our catalogue from the standard image.  The GLEAM number counts at 162\,MHz and source counts in our standard image are expected to be comparable, since both are dominated by the same extragalactic source population. Our field is close to the Galactic South pole, while the published GLEAM catalogue has a Galactic latitude limit of 10$^\circ$.

\begin{center}
\begin{figure}
\includegraphics[scale=0.33, angle=0]{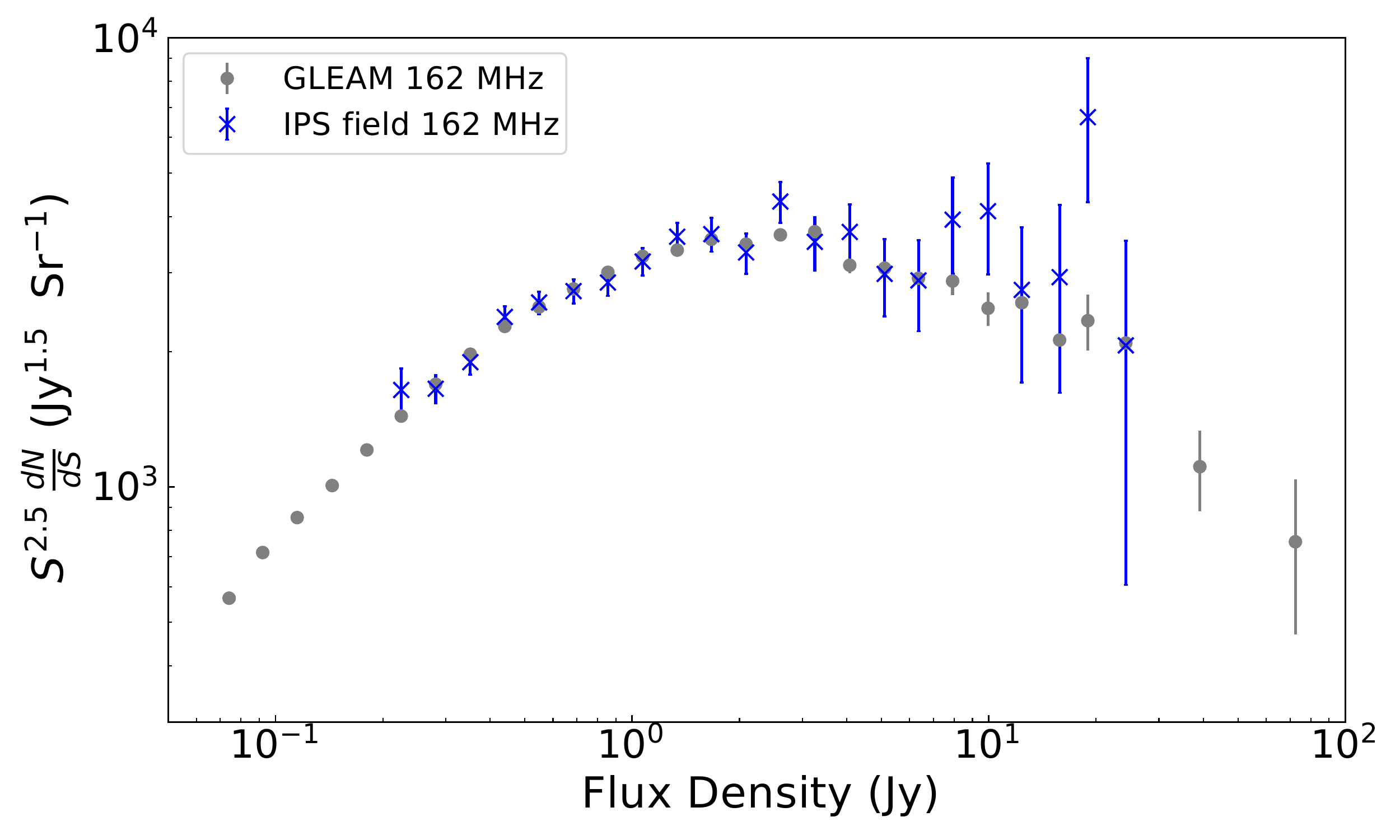}
\caption{We compare the differential source counts, here plotted weighted to Euclidean source counts, for our total source sample (blue triangles) against the 154\,MHz GLEAM counts extrapolated to 162\,MHz with $\alpha$=-0.8 (grey filled circles). This agreement between the two provides evidence of the robustness of our analysis technique.}
\label{Fig:compare_srcCount}
\end{figure}
\end{center}

We plot the counts in Figure \ref{Fig:compare_srcCount} and compare them against the source counts for GLEAM at 162 MHz. 
The source count statistics of our field show very good agreement with the GLEAM counts.  The excellent match between the counts of our parent sample and the GLEAM catalogue shows that our technique of estimating the counts from sources detected in a variable sensitivity image is robust.
We note that there are deviations between the counts at the high and low flux density limits of the survey; at the limit of sensitivity, incompleteness and small number statistics cause deviations in the number counts from the more sensitive GLEAM survey number counts. In the highest flux density bins the small number of sources results in a larger statistical uncertainty.

\subsection{Source counts of compact sources}
\label{Sec:noCountCompactSrc}

In paper\,II we presented a catalogue of compact sources. The compact flux densities of the sources were not included in the catalogue explicitly. These flux densities are obtained from the primary beam corrected and absolute flux density corrected variability image, with appropriate background correction discussed in section 3.2.1 of Paper II.

\subsubsection{Estimation of the sensitivity to compact sources}
As with the Total Population source counts, we wish to determine the (5-$\sigma$) detection limit for a compact source for each point in our image. As noted above, due to the dependence of the scintillation on distance from the Sun, and the unusual statistics of our variability image, this requires steps additional to those applied to the standard image.

As before, Aegean \citep{Hancock2012} was used to estimate the positive background bias of the variability image, and its RMS.
All three images (the variability image and its background and RMS) were then scaled to account for primary beam effects.
The background and the RMS of each pixel, termed $\mu$ and $\sigma$ in Paper I, can then be used to determine the scintillating flux density which would be detected at the 5-$\sigma$ level, $dS_{min}$ (see equation 4 in paper I):
\begin{equation}
  dS_{\rm min} = \sqrt{(5\sigma + \mu)^2 - ( \mu )^2 } .
  \label{Eqn:bkgrd_removal_noise}
\end{equation}

The compact flux density of a scintillating source depends on its scintillation index, which in turn depends on its distance from the Sun.
Thus the final step in the formation of our compact source sensitivity image is to divide $dS_{min}$ by the expected scintillation index at the elongation of each pixel from the Sun (see section 4.5 of Paper I and section 3.3 of Paper II).

\begin{center}
\begin{figure}
\includegraphics[scale=0.55, angle=0]{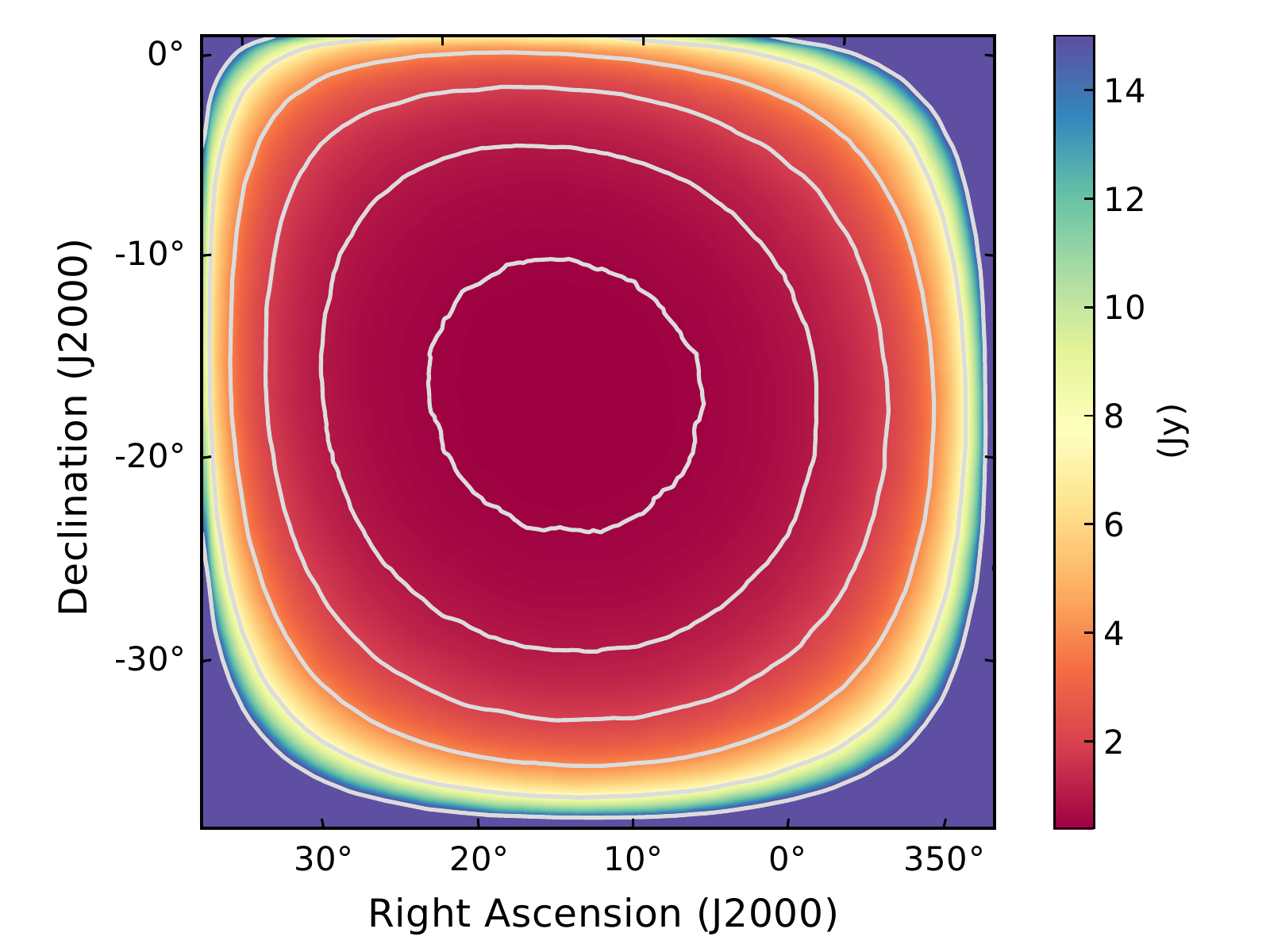}
\caption{The region of sky encompassing regions 5\,$\sigma$ sensitivity as a function flux density in our variability image. Contours are drawn at 0.5, 1, 2, 4, 8 and 16 Jy. }
\label{Fig:sensitivity_area}
\end{figure}
\end{center}

\begin{figure}
	\includegraphics[scale=0.33]{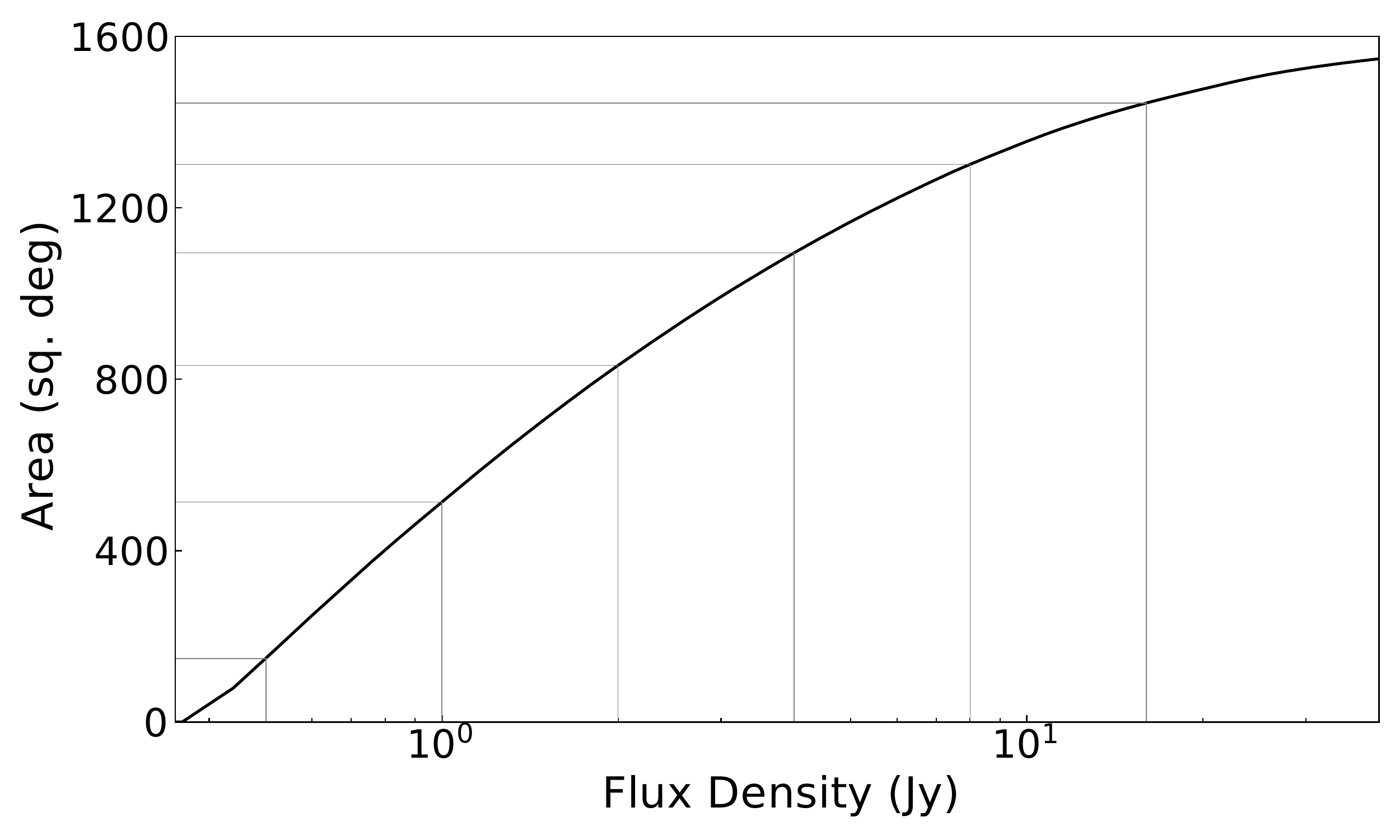}
    \caption{Plot of the profile of available areas in our sensitivity image at different flux densities. Grey vertical lines are drawn at same flux density values as the contours in Fig.\,\,\ref{Fig:sensitivity_area} (0.5, 1, 2, 4, 8 and 16 Jy). The grey horizontal lines show the corresponding areas of sensitivity in the image, and the sky, at those flux density values.}
    \label{Fig:sensitivityArea}
\end{figure}

Figure \ref{Fig:sensitivity_area} shows the resulting sensitivity image. The effect of solar elongation on sensitivity is seen as a protraction of contours towards the upper left corner (the Sun was located towards the upper left part outside the image during this observation). Fig.\,\,\ref{Fig:sensitivityArea} plots the area corresponding to each compact flux density limit.

\subsubsection{Correction for Eddington bias}
For a source population with a distribution that declines to higher flux density, the number of sources just above a survey detection threshold is artificially raised by the larger number of sources whose flux density are upscattered due to noise. This is the Eddington bias \citep{Eddington_1913MNRAS..73..359E}, since re-discovered many times and discussed in various contexts \citep[e.g.][]{Jauncey_1967Natur.216..877J}. 

For IPS observations, variations in the solar wind speed across the field of view introduce an additional source of uncertainty in converting from scintillating flux density to compact flux density. 
This noise is different from the system noise case in that it is multiplicative: the fractional error is the same regardless of the signal to noise of the source.
For our field, we estimate this uncertainty to be approximately $\sim$20\%, since this is the scatter in NSI about 1.0 for flat spectrum (i.e. likely very compact) sources (see Paper II, Fig. 4).

We conducted a Monte-Carlo simulation to estimate and remove the Eddington bias in our data due to these two effects combined. With no previous number counts of compact sources to guide our approach at the frequency and flux densities we probe, we used the slope of the GLEAM sources count distribution close to the detection limit of our scintillating sources ($dN/dS \propto S^{-1.9}$).  

We applied a correction factor, based on our simulations with 10\,000 realisations, to the count in each bin to account for this Eddington bias.

\section{Result and Discussion}
\label{Sec:Discussion}
We present two distinctly different source counts: (i) the counts of any flux density in compact structures in all the Parent Population, which we term the \emph{counts of sources with compact structure}, and (ii) the \emph{point source counts} where we consider only those sources which are unresolved to our IPS technique. Both have precedents in the literature and both can be useful as described below.  

In constructing the counts of sources with compact structure we use all sources which show an IPS detection, and the flux density used is the compact flux density. We caution that these counts do not lend themselves to straightforward comparison with standard source counts because for sources that are not entirely point-like, the compact flux represents only a part of the total source flux density (note that this could be a distinct component or component, such as a core or hotspot(s), or a single source which is slightly extended).  For instance, a source with compact flux density of 0.4\,Jy may equally be associated with 100\% compact 0.4\,Jy source or with a 1\% compact 40\,Jy source. Furthermore, at the 5$\sigma$ noise limit of 0.4\,Jy a 40\,Jy source would be included in the compact counts even if it were only 1\% compact, whereas a 0.4\,Jy source would be included only if it were entirely compact. 

For the point source counts, we use those 93 sources whose entire flux densities arise from a sub-arcsecond compact region, as identified by the criterion NSI $\geq$ 0.9. By restricting the analysis to this specific population where the compact flux density is the same as the total flux density, we avoid the complications that apply to the counts of sources with compact structure.

We show in Fig.~\ref{Fig:CompactFlux_srcCount} the source count distribution of the sources with compact structure while Fig.~\ref{Fig:CompactSources_srcCount} shows the corresponding distribution for the point sources.
The counts for the point source distribution are listed in Table \ref{Tab:StrongScint}. The 162 MHz GLEAM number counts are plotted in grey for comparison in both figures.

\begin{table*}
\small
\resizebox{\textwidth}{!}{
\begin{tabular}{@{\extracolsep{2pt}} cccccccccc}
\hline
&	&	\multicolumn{4}{c}{\textbf{Sources with compact structure}} & \multicolumn{4}{c}{\textbf{Point sources}} \\
\cline{3-6}
\cline{7-10}\\
\textbf{Bin} & \textbf{Bin} &	 	\textbf{Raw}  &	\textbf{Median S} & \textbf{Counts}  & \textbf{Err} &	 	\textbf{Raw}  &	\textbf{Median S} & \textbf{Counts}  & \textbf{Err}\\
\textbf{bottom} & \textbf{top}	&  \textbf{Counts}	  & \textbf{(Jy)} &	\multicolumn{2}{c}{\textbf{(Jy$^{1.5}$sr$^{-1}$)}}&  	\textbf{Counts}  & \textbf{(Jy)} &	\multicolumn{2}{c}{\textbf{(Jy$^{1.5}$sr$^{-1}$)}} \\
(1) & (2) &	(3)	& 	(4) & (5) &	(6)	& 	(7) & (8) & (9) &(10)\\
\hline

0.39	&	0.49	&	15	&	0.47	&	733	&	189	&	5	&	0.48	&	208	&	93\\
0.49	&	0.62	&	25	&	0.58	&	705	&	141	&	11	&	0.57	&	302	&	91\\
0.62	&	0.77	&	39	&	0.69	&	911	&	146	&	14	&	0.69	&	328	&	88\\
0.77	&	0.96	&	39	&	0.85	&	915	&	147	&	12	&	0.81	&	259	&	75\\
0.96	&	1.20	&	23	&	1.10	&	643	&	134	&	10	&	1.08	&	273	&	86\\
1.20	&	1.50	&	33	&	1.29	&	975	&	170	&	10	&	1.25	&	280	&	88\\
1.50	&	1.87	&	20	&	1.69	&	790	&	177	&	7	&	1.74	&	293	&	111\\
1.87	&	2.34	&	15	&	2.17	&	771	&	199	&	8	&	2.23	&	438	&	155\\
2.34	&	2.93	&	14	&	2.63	&	822	&	220	&	5	&	2.62	&	293	&	131\\
2.93	&	3.67	&	4	&	3.33	&	319	&	160	&	1	&	3.27	&	76	&	76\\
3.67	&	4.58	&	7	&	3.92	&	581	&	220	&	5	&	3.92	&	414	&	185\\
4.58	&	5.72	&	3	&	5.36	&	397	&	229	&	0	&	-	&	-	&	-\\
5.72	&	7.15	&	6	&	5.93	&	713	&	291	&	3	&	6.3	&	412	&	238\\
7.15	&	8.94	&	2	&	7.84	&	332	&	235	&	1	&	7.97	&	173	&	173\\
8.94	&	11.14	&	0	&	-	&	-	&	-	&	0	&	-	&	-	&	-\\
11.14	&	14.02	&	1	&	12.89	&	193	&	193	&	1	&	12.89	&	193	&	193\\
14.02	&	17.48	&	1	&	14.35	&	140	&	140	&	0	&	-	&	-	&	-\\
\hline

\end{tabular}
}
\caption{Table of Euclidean-weighted differential counts sources with compact structure, and point sources. Median flux density of all sources in the bin used to Eucledean weight the number counts are presented in ``Median S" columns. Figures \ref{Fig:CompactFlux_srcCount} and \ref{Fig:CompactSources_srcCount} compare these counts against the GLEAM number counts of entire population of extragalactic sources at 162 MHz.
The columns are as follows: $(1)$ Lower end of bin range (Jy).; $(2)$ Higher end of the bins range (Jy); $(3)\, \&\, (7)$ Number of sources in the bin;
$(4)\, \&\, (8)$ The median flux density of all sources in the bin (Jy) used to Euclidean-weight the counts; $(5)\, \&\, (9)$ The differential number counts, normalised to the value expected for a Euclidean source counts distribution, ie. multiplied by S$^{2.5}$;  $(6)\, \&\, (10)$ Uncertainty in counts.  
}
\label{Tab:StrongScint}
\end{table*}

\subsection{Counts of sources with compact structure}

\begin{figure*}
	\includegraphics[scale=0.7]{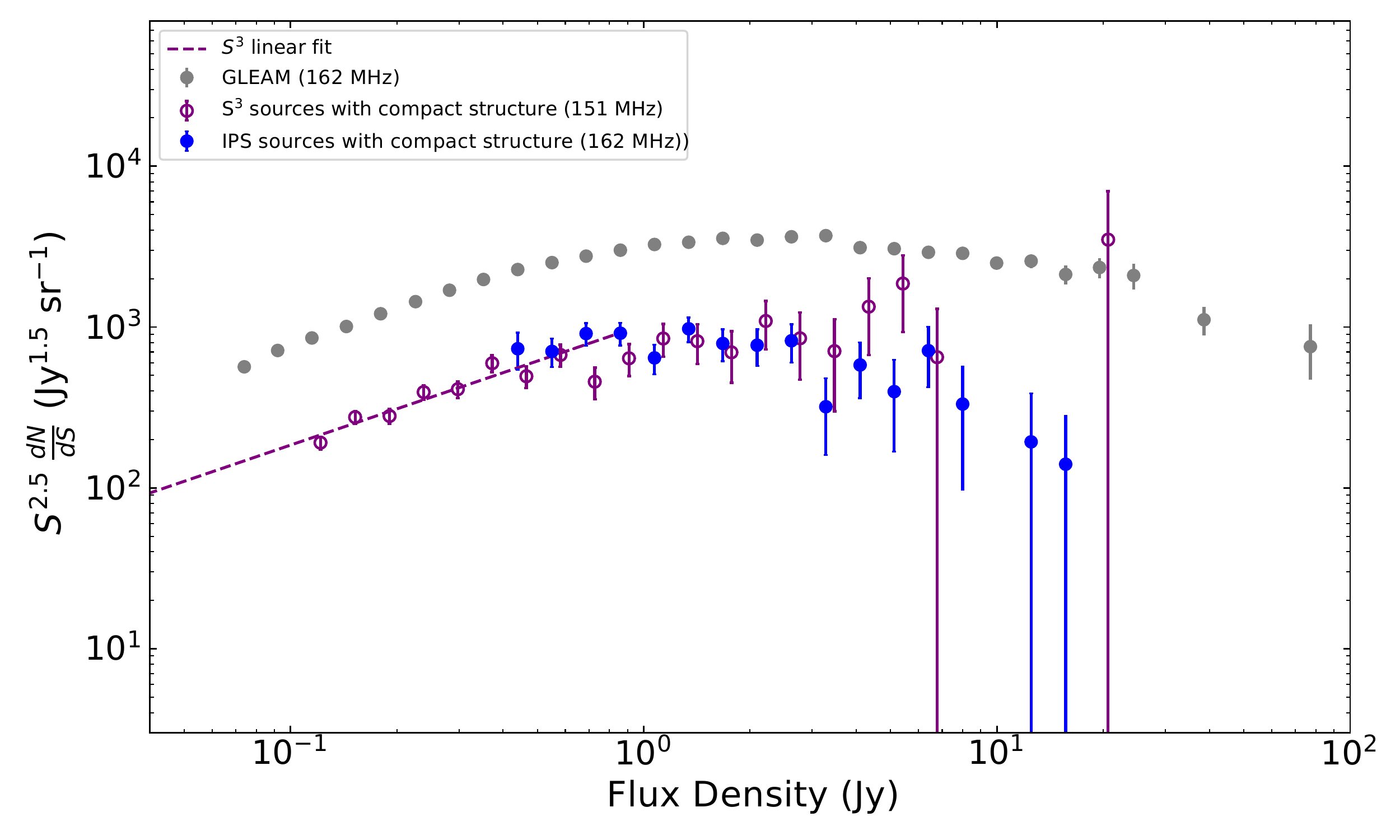}
    \caption{We present counts of sources with compact structure at 162 MHz (blue filled circles) and compare them against the overall GLEAM number counts (grey filled circles). In many cases only a fraction of the source is counted. The GLEAM counts are translated to 162 MHz for direct comparison. The uncertainty in the counts is estimated from the Poisson error based on the raw count value. Counts of sources with compact structure with our technique for the S$^3$ simulation at 151 MHz (see text) are presented (purple open circles), truncated at 100 mJy. Purple dotted line represents a linear fit ($\frac{dN}{dS}\,=1032.1$ S$^{-1.75}$ Jy$^{-1}$sr$^{-1}$) to the S$^3$ simulation compact detections, below our detection limit.}
    \label{Fig:CompactFlux_srcCount}
\end{figure*}

The counts of sources with compact structure presented in Fig. \ref{Fig:CompactFlux_srcCount} follow the overall population above $\sim$3\,Jy, within the range of uncertainties, then appear Euclidean down to our sensitivity limit of 0.4\,Jy. Our results of these counts can be compared to the work by \cite{Artyukh_1996ARep...40..601A}. They conducted a similar survey of radio sources exhibiting interplanetary scintillation at 102\,MHz. The IPS survey was made using the Large Phased Array (LPA) of the Lebedev Institute of Physics in Pushchino. The LPA has a very large collecting area (30\,000 m$^2$ at the zenith) and hence very high instantaneous sensitivity to IPS but has much lower angular resolution than the MWA, so is seriously influenced by source confusion. To overcome this limitation they used the 7C radio source survey \citep{McGilchrist_1990MNRAS.246..110M} to identify the scintillating sources and to obtain a value for the total (non-scintillating) flux density. Despite this the scintillating sources are themselves still affected by confusion in the Pushchino survey and source identifications are often ambiguous. Our MWA observations with $>$2550 beam areas per scintillating source have negligible confusion and the total intensity can be un-ambiguously determined from the same IPS observations. Our result of counts of compact detections  are in excellent agreement with their Figure 1 above a flux density of 0.5\,Jy, which is their stated completeness limit.  We even see the slight (but probably not significant) excess in the range 0.5 to 1.0\,Jy.  However, we do not observe the sharp drop in the differential counts of scintillating sources below 0.6\,Jy reported in their paper. While we do not have sufficient sensitivity to probe counts at the faintest level they reach, we suggest that they have not been able to fully correct for completeness and confusion at these faint levels. 

\cite{Wilman_2008MNRAS.388.1335W} have constructed a simulation of 20\degr$\times$20\degr of extragalactic radio continuum sky, as part of the SKA Simulated Skies (S$^3$) project\footnote{\url{http://s-cubed.physics.ox.ac.uk/}}. The output of the simulation consists of various ``Galaxy'' types (FRI, FRII, Quasars and star-forming galaxies) made up of sets of ``components'' (cores, lobes, hotspots and disks) appropriate to each galaxy type. The flux density of each component at a variety of frequencies including 151\,MHz ($S_{151}$) is provided, as well as each component's morphology (major axis $a$, minor axis $b$ and position angle). We derived the counts of compact objects predicted by this simulation as follows. First, all components for all galaxies with total flux density greater than 100\,mJy were selected. The compact flux density of each component (as would be measured by IPS) was estimated to be

\begin{equation}
	S_{\rm IPS} = S_{151} \times \left\{ \begin{array}{ll} 
    \frac{\theta_F}{ \sqrt{a^2+b^2} }, & \sqrt{a^2+b^2}> \theta_F \\
    1, & \hbox{otherwise}. \\
    \end{array} \right. 
	\label{Eqn:CompactFlux}
\end{equation}

where the Fresnel scale, $\theta_F$, is assumed to be 0.3\arcsec (see Section 4.1 of Paper II for further details).
Next, each group of components that lay within $\theta_F$ of each other were blended into a single component by summing their IPS flux densities (as is appropriate for \emph{coherent} combination of IPS variability).  Then, each group of components that lay within 3\arcmin (the inteferometric resolution of the MWA) were blended into a single component by summing their IPS flux densities in quadrature (as is appropriate for \emph{incoherent} combination of IPS variability). The resulting sources with a flux density > 100\,mJy were preserved. 
The difference in frequencies (their 151\,MHz vs. our 162\,MHz) is assumed to be negligible.

We can now compare the resulting source counts of this simulation against our counts of sources with compact structure (see Fig.~\ref{Fig:CompactFlux_srcCount}).   As noted by \cite{Wilman_2008MNRAS.388.1335W} the (S$^3$) simulations should be used with caution outside the range of parameters which are constrained by their current (i.e. 2008) knowledge but our IPS observation can be used to test aspects of the model using new information not available in 2008 and as appropriate can be used to improve the simulation model. In our flux density range we need only consider the radio loud AGN components of the model which includes the cores, lobes and hot spots.  Our data shows a remarkable agreement in the density of sources with compact structure and reasonable agreement for the slope of the source counts for these sources. Inspection of the simulated component list at 162 MHz shows that our counts of all sources with compact structure is dominated by hotspots so we can explore this aspect of the model. The agreement in density confirms the use of old hot spot data from \cite{jenkins1977MNRAS.180..219J} which sets the fraction of the total flux density in the hot spots.  The fact that the slope of the simulated counts for scintillating components is the same as the counts for total flux density is no surprise since \cite{Wilman_2008MNRAS.388.1335W} assume that hot spots have the same luminosity function and spectral shape as the lobes.  The actual agreement with our observed slope validates this quite reasonable assumption.  

\subsection{Counts of point sources}
\label{Sec:CountsPointSrc}
\begin{figure*}
	\includegraphics[scale=0.7]{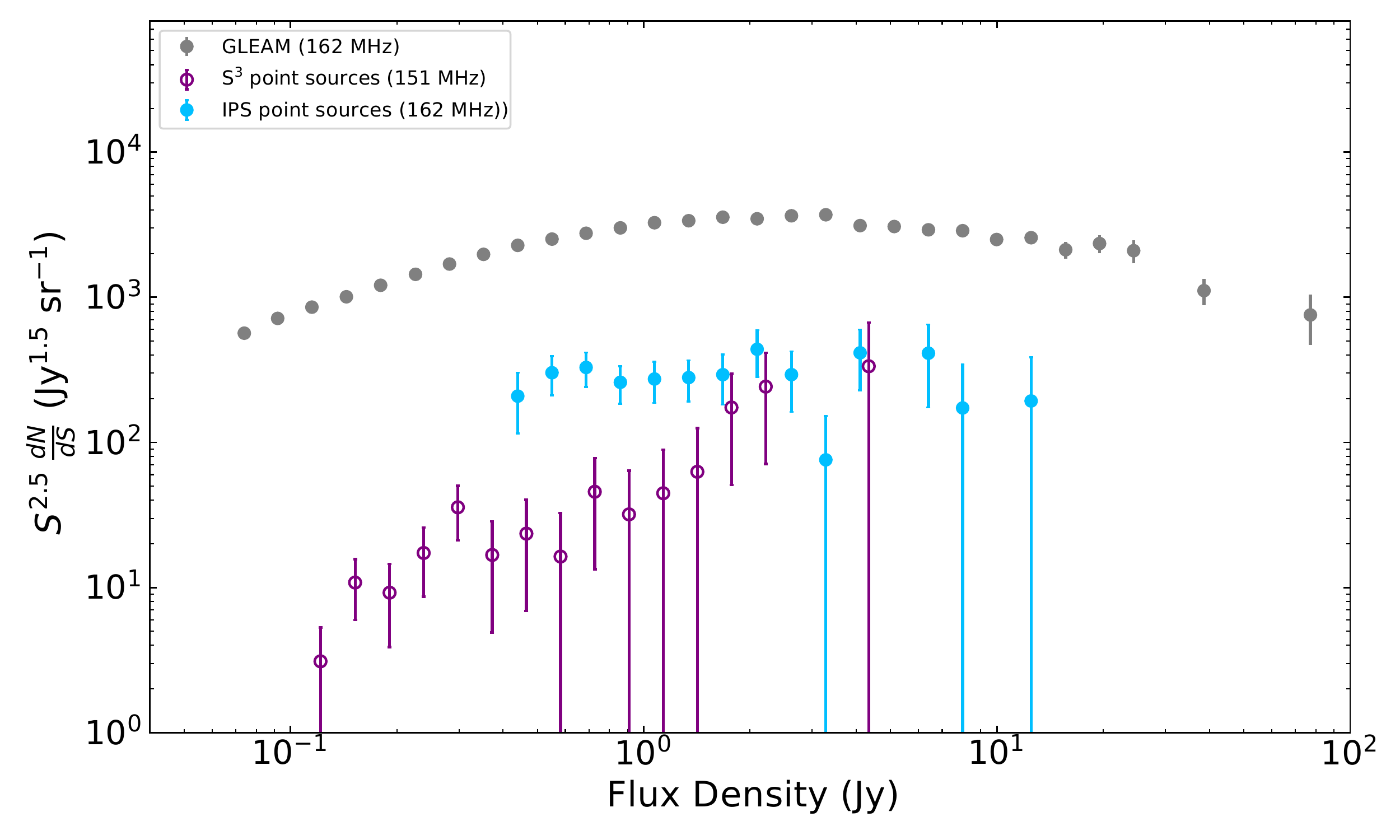}
    \caption{We present Euclidean-weighted differential number counts of point sources with blue filled circles. The flux density for these sources are expected to be confined to 0.3 arcsecond at 162 MHz. We plot the number counts of overall GLEAM population in grey filled circles for comparison. The uncertainty in the source counts is estimated from the Poisson error based on the raw count value. Counts of point sources obtained from the S$^3$ simulations at 151\,MHz are shown with purple open circles.}
    \label{Fig:CompactSources_srcCount}
\end{figure*}

For the reasons given above the source counts of all 247 objects showing scintillation presented in the previous section do not lend themselves to direct cosmological interpretation. We now consider the more readily interpretable counts of point sources which are presented in Fig. \ref{Fig:CompactSources_srcCount}.

\cite{Readhead_1975MNRAS.170..393R} selected quasars from the 3C catalogue with flux density > 2 Jy and used IPS to identify sources with fractional flux from compact components $\geq$ 0.40. They interpreted the resulting source counts to be evidence for strong evolution of quasars. Our sky coverage is much smaller than the total 3C catalogue, limiting our numbers at the high flux density end. By contrast, we probe an order of magnitude lower in flux density and restrict the fractional flux for compact components to $\geq$ 0.90, selecting only sources smaller than 0.3 arcseconds.

The counts obtained for our point source population, shown in Fig. \ref{Fig:CompactSources_srcCount}, show a Euclidean profile (ie flat in this normalized plot) across the entire flux density range that we are sensitive to, over almost two orders of magnitude. The GLEAM source counts deviate from Euclidean below $\sim$ 1 Jy. A difference in the two distributions would be consistent with the understanding that a) the extended radio galaxies and the compact point source populations (eg. AGNs) have different lifespans so they must evolve differently, b) that their luminosity functions must be different, or c) a combination of both.

Also plotted in Fig. \ref{Fig:CompactSources_srcCount} are the source counts from the S$^3$ simulations.
The NSI for each simulated source was determined by counting the flux density of \emph{all} components that lay within an MWA beam of a compact source, allowing the ratio of compact vs. total flux density (i.e. the NSI) to be computed.
Looking at those that would be detected by the MWA as point sources, we see that, in contrast to the compact component source counts, the simulation is not able to reproduce our observed point source counts.
With such a complex simulation, determining which input parameters might be adjusted in order to produce output more consistent with our results is not straightforward, especially with only one realisation of the simulation which covers less than a steradian, and so we do not attempt this here.
However we note that the combination of compact morphologies from IPS, along with detailed spectral information from GLEAM will provide strong constraints for testing such models.

One important result from paper II was that $\sim$ 40\% of the compact source population was composed of peaked spectrum objects, and interestingly the S$^3$ simulations do define 19/34 galaxies that we classify as compact as `GPS' AGN. 
\cite{Callingham_2017ApJ...836..174C} identified over 1400 peaked spectrum sources in the GLEAM survey, including sources with peaks outside the GLEAM frequencies. Peaked spectrum sources (eg. the gigahertz peaked-spectrum (GPS) sources) show morphologies similar to FR II sources but are confined to <10 kpc in linear size, and are generally considered to be younger stages of radio galaxies \citep[eg.][]{O'Dea1998}. Using their catalogue, we found 34 peaked-spectrum sources in our point source population. One source (GLEAM J013243-165444) that is identified as a peaked spectrum object was found to be a blazar in paper II, and not a peaked spectrum source in the traditional sense. We removed this source from our list, leaving 33 peaked spectrum sources. 

\begin{figure}
	\includegraphics[scale=0.43]{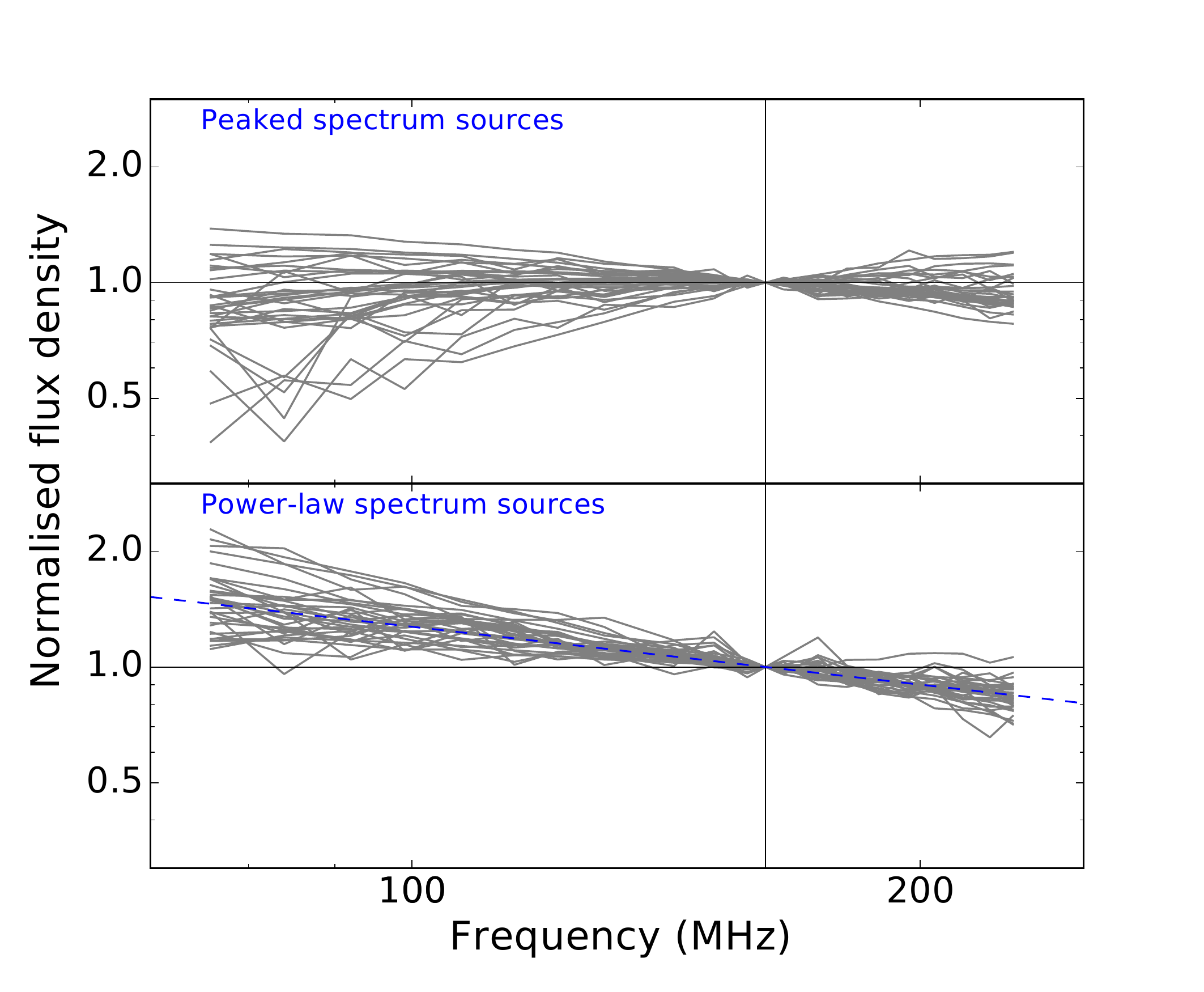}
    \caption{The normalised SEDs of the sub-populations of point sources in our sample, made using GLEAM flux densities normalised by their estimated flux density at 162 MHz (vertical line). The top panel shows the peaked spectrum sources and the bottom panel the power-law spectrum sources. Dotted blue line in bottom panel denotes the $\alpha$=-0.5 limit used to identify compact steep-spectrum sources.}
    \label{Fig:moustache_plot}
\end{figure}

The remaining 60 sources in our point source population show power-law spectra with spectral indices on a continuum, with 18  of them showing flat-spectrum ($\alpha_{GLEAM}\,\geq$ -0.5) and the remainder steep ($\alpha_{GLEAM}\,<$ -0.5) as seen in Fig. \ref{Fig:moustache_plot}, and Table \ref{Tab:NoSrc_Spectra}. Note that there is no obvious bimodality in the distribution of spectral indices; the classification is purely conventional. The flat-spectrum sub-population of point sources are the beamed compact cores of radio galaxies. Only one source (the above mentioned GLEAM J013243-165444) shows Fermi gamma-ray association. Here, we note that the flat- or steep-spectrum hot-spots on radio galaxies only form a minor part (<2\%, based on work by \cite{Readhead_1976MNRAS.176..571R}) of this population as they mostly show NSI lower than 0.9 by virtue of their being embedded in extended radio lobes or having two or more compact components with their IPS flux adding incoherently to lower their NSI. 

The different spectral properties of sub-populations that form our point source population are summarised in Table 2. 

In Fig. \ref{Fig:moustache_plot} we plot the low frequency SEDs of individual peaked and non-peaked spectrum sources separately. Their spectral properties and their SED inform us that the counts seen in Fig. \ref{Fig:CompactSources_srcCount} are a composite of the source counts of these two different sub-populations, and that their different evolution and luminosity function profiles pose challenge to a direct interpretation of Fig. \ref{Fig:CompactSources_srcCount}.

At high frequencies, the synchrotron self-absorbed cores of AGNs show flat spectra and dominate the total radio source population, while a class of compact objects with steep spectra form a small fraction of the Total Population \citep{Chhetri2013}. These compact steep-spectrum (CSS) sources are considered to be at more evolved stages in radio galaxy evolution than the GPS sources. With the hypothesis that the steep-spectrum compact objects in our subsample are CSS objects, with their spectral peaks below frequencies covered by GLEAM, we combined them with the peaked source population to form a single sub-population. With this, the GPS/CSS objects become the single most dominant group, comprising 81\% of the total low frequency sub arcsecond population, while the flat-spectrum objects form only a minority (19\%). 

\begin{table}
\centering
\resizebox{\columnwidth}{!}
{
\begin{tabular}{lccc}
\hline
\textbf{Sub population}	&	\textbf{NSI}		&	 	\textbf{Number}	&\textbf{Median Spectral} \\
				&	\textbf{criteria}	& 	\textbf{of sources}	&\textbf{index} \\
\hline 
Parent population$^*$			&		all			&  247	&	-0.62\\    
Point sources$^*$				&	$\geq$0.90		&	93	&	-0.49\\    
Peaked-spectrum sources$^*$ 	&	$\geq$0.90		&	33	&	-0.23\\    
CSS sources		&	$\geq$0.90		&	42	&	-0.66\\    
Flat-spectrum sources		&	$\geq$0.90		&	18	&	-0.39\\    
\hline
\end{tabular}
}
\caption{Table lists the NSI criteria, number of sources and median spectral indices for different sub-populations. `*' represents sub-populations that contain peaked-spectrum sources -- sources that do not have a single representative spectral index. For these sub-populations median spectral indices are estimated using GLEAM flux densities at 158 and 166\,MHz, closest frequencies to our observation. For the remaining sub-populations showing power-law, medians estimates are based on spectral indices in the catalogue in paper II. }
\label{Tab:NoSrc_Spectra}
\end{table}

\begin{figure*}
	\includegraphics[scale=0.7]{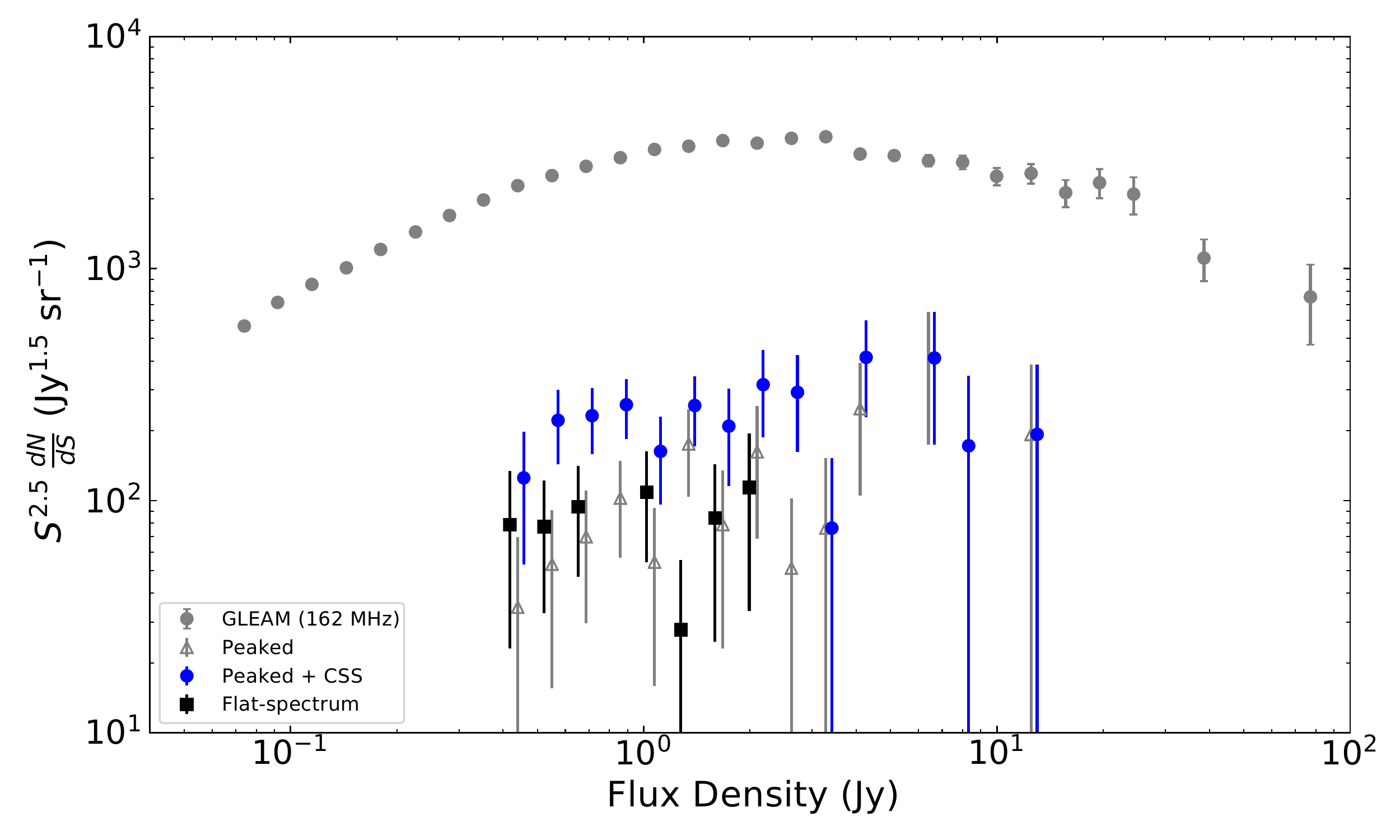}
    \caption{We present and compare the Euclidean-weighted differential number counts for the different sub-populations that make up the point source population. These are the peaked-spectrum source population (grey open triangles), the peaked spectrum sources combined with the compact steep-spectrum sources (blue filled circles), and the flat-spectrum sources (black filled squares). For clarity, the X-positions of filled circles and squares have been shifted towards the right and left by 5\% of their values respectively. Counts of overall GLEAM are plotted with grey filled circles for reference. The uncertainty in the source counts is estimated from the Poisson error based on the raw count value.}
    \label{Fig:compare_SubPopn_Counts}
\end{figure*}

We made counts of the different sub-populations identified above and compare them in Figure \ref{Fig:compare_SubPopn_Counts}. We tested against the hypothesis that the counts of different sub-populations are consistent to within a constant multiplicative factor of the GLEAM counts and other counts of interest (ie. counts of sub-populations = f $\times$ GLEAM counts), where the fraction ``f" is the best-fitting scaling parameter. Table \ref{Tab:chiSq_stats} lists the reduced $\chi^2$ values and the scaling factors used for different sub-populations. The reduced $\chi^2$ were calculated as:

\begin{equation}
	 \Delta\,\chi^2 = \frac{1}{(N-1)} \sum_{i=1}^{N}  \frac{(\mu_{1_i} - f\,*\, \mu_{2_i})^2}{(\sigma_{1_i}^2 +\sigma_{2_i}^2)}
	\label{Eqn:redChiSq}
\end{equation}
where, N is the number of flux density bins, $\mu_{1}, \sigma_{1}$ are the bin values and uncertainties for the population under consideration, and $\mu_{2}, \sigma_{2}$ are the bin values and uncertainties for the model.

We did not undertake a more rigorous comparison of the source count distributions using, for instance, a two-sided Kolomogorov-Smirnov test, because it is already clear from our crude comparison of the binned data using the reduced $\chi^2$ metric that the errors in various distributions are too large to identify any points of difference between the distributions. We expect this picture to change with better statistics achievable with our future widefield IPS survey. 

Currently, an improved understanding may be gained from the counts of peaked spectrum sources made using \cite{Callingham_2017ApJ...836..174C} data with appropriate completeness corrections to lower flux densities. Here as well, future widefield IPS survey will provide complementary information by identifying CSS sources to complete the picture.

\begin{table*}
\resizebox{\textwidth}{!}{
\begin{tabular}{@{\extracolsep{1pt}}lccccccccc}
\hline
\textbf{} &  \multicolumn{3}{c}{\textbf{GLEAM}} & \multicolumn{3}{c}{\textbf{Point sources}} & \multicolumn{3}{c}{\textbf{Peaked+CSS sources}}
\\
\cline{2-4}
\cline{5-7}
\cline{8-10}

\textbf{Sub-Population} & \textbf{f} & \textbf{DoF}  & \textbf{$\Delta\,\chi^2$} & \textbf{f} & \textbf{DoF}  & \textbf{$\Delta\,\chi^2$} & \textbf{f} & \textbf{DoF}  & \textbf{$\Delta\,\chi^2$}   \\

(1) & (2) &	(3)	& 	(4) & (5) &	(6) & (7) & (8) & (9) & (10)\\
\hline 
Point sources & 0.080 & 14 & 1.06 & - & - & - & - & - & - \\
Peaked-spectrum sources & 0.024 & 13 & 0.68 & - & - & - & - & - & - \\
Peaked + CSS sources & 0.065 & 14 & 0.78 & 0.81 & 14 & 0.11 & - & - & - \\
Flat-spectrum sources & 0.020 & 8 & 0.73 & - & - & - & 0.34 & 7 & 0.58 \\

\hline
\end{tabular}
}
\caption{The result of reduced-$\chi^2$ test performed to test against the hypothesis that counts of different sub-populations are consistent to a constant multiplicative factor of other sub-populations. The fraction ``f" is the best-fitting scaling parameter used (see text), ``DoF" is the degree of freedom and ``$\Delta\,\chi^2$" is the reduced-$\chi^2$ value.}

\label{Tab:chiSq_stats}
\end{table*}

\subsection{Density of calibrators}
\label{Sec:implicationsForSKA}
Knowledge of the compact source counts is also important for the design and use of long-baseline interferometers, including the SKA\_LOW which will reach arcsecond resolution on its longest baselines.
\cite{Moldon_2015A&A...574A..73M} and subsequently \cite{Jackson_2016A&A...595A..86J} have conducted a search for compact sources that are suitable calibrators for international baselines of the LOFAR telescope with correlated flux density of $\gtrsim$50--100 mJy at frequencies around 110--190 MHz, on scales of a few hundred milliarcseconds. They find around 1 suitable calibrator source per square degree on 200-300 km international baselines of LOFAR, which drops to 0.5 sources per square degrees on 200-600 km baselines. 

\citeauthor{Moldon_2015A&A...574A..73M} also address the problem of selecting likely compact calibrators \textit{a priori}. At higher frequencies this is done by selecting flat-spectrum sources, which is a very well-justified approach, since these are extremely likely to be very compact \citep{Chhetri2013}. \citeauthor{Moldon_2015A&A...574A..73M} note that spectral index at higher frequencies is a poor predictor of compactness at low frequencies; however they find that flat spectra at low frequencies is much more likely to select satisfactory primary calibrators. Fig.~\ref{Fig:compare_SubPopn_Counts} clearly shows that by selecting peaked sources \emph{as well as} flat-spectrum sources, the number of high-quality compact calibrators that will be selected is almost doubled in the flux density range 0.4--2\,Jy. Additionally, significant numbers of calibrators at even higher flux densities will be selected. Furthermore, we showed in Paper II that all peaked sources in the Total Population were compact, so the use of peaked sources in addition to flat-spectrum sources does not compromise reliability.

The calibration of SKA\_LOW is entirely dependent on having a grid of calibrators in the field of view (FoV) dense enough to provide the direction dependent calibration of multiple ionospherically limited patches. To this effect, we can provide information at two levels. The density of point sources on baselines out to about 350\,km at 162MHz (< 1$\arcsec$ in size)  is  0.2 per square degree at 0.4\,Jy (this is approximately an order of magnitude higher than that estimated by the S$^3$ model in Fig. \ref{Fig:CompactSources_srcCount}). However if more complex sources can be modeled in the calibration strategy we can use all scintillating sources with a density of 0.7 per square degree and these will have at least some visibility on baselines out to 350 \,km.

The S$^3$ model in Fig. \ref{Fig:CompactFlux_srcCount} indicates that the density of compact components follows a trend similar to the overall low radio frequency source population below $\sim$ 0.4\,Jy. Although these simulations cannot exactly reproduce our results at higher flux densities, in the absense of real measurements we use the simulations as a guide to extrapolate down another order of magnitude and estimate a density of compact source density of 4.6 sources/deg.$^{2}$ at 40\,mJy, a flux density scale useful for SKA\_LOW calibrators.

\begin{table*}
\resizebox{\textwidth}{!}{
\begin{tabular}{@{\extracolsep{4pt}}cccccccccccccc}
\hline
\textbf{Bin} & \textbf{Bin} &   \multicolumn{4}{c}{\textbf{Peaked sources}} & \multicolumn{4}{c}{\textbf{Peaked + CSS sources}} & \multicolumn{4}{c}{\textbf{Flat-spectrum sources}}
\\
\cline{3-6}
\cline{7-10}
\cline{11-14}

\textbf{bottom} & \textbf{top} & \textbf{Raw}  & \textbf{Median S} &	\textbf{Counts} & \textbf{Err} & \textbf{Raw}   & \textbf{Median S} & \textbf{Counts} & \textbf{Err} & \textbf{Raw} & \textbf{Median S}  & \textbf{Counts} & \textbf{Err}\\

\textbf{(Jy)} & \textbf{(Jy)}	& \textbf{counts} & \textbf{(Jy)} & \multicolumn{2}{c}{ \textbf{($Jy^{1.5}Sr^{-1}$)}}& 
\textbf{counts} & \textbf{(Jy)} & \multicolumn{2}{c}{ \textbf{($Jy^{1.5}Sr^{-1}$)} } & \textbf{counts} & \textbf{(Jy)} & \multicolumn{2}{c}{ \textbf{($Jy^{1.5}Sr^{-1}$)} }\\

(1) & (2) &	(3)	& 	(4) & (5) &	(6) & (7) & (8) & (9) & (10) & (11) & (12) & (13) & (14)\\
\hline 

0.39 & 0.49 & 1 & 0.48 & 35 & 35 & 3 & 0.48 & 125 & 72 & 2 & 0.47 & 79 & 56\\
0.49 & 0.62 & 2 & 0.57 & 53 & 38 & 8 & 0.57 & 222 & 78 & 3 & 0.58 & 77 & 45\\
0.62 & 0.77 & 3 & 0.67 & 70 & 40 & 10 & 0.69 & 232 & 74 & 4 & 0.68 & 94 & 47\\
0.77 & 0.96 & 5 & 0.79 & 103 & 46 & 12 & 0.81 & 259 & 75 & 0 & - & - & -\\
0.96 & 1.20 & 2 & 1.08 & 54 & 38 & 6 & 1.08 & 163 & 67 & 4 & 1.07 & 109 & 54\\
1.20 & 1.50 & 6 & 1.27 & 176 & 72 & 9 & 1.26 & 257 & 86 & 1 & 1.23 & 28 & 28\\
1.50 & 1.87 & 2 & 1.69 & 79 & 56 & 5 & 1.74 & 209 & 94 & 2 & 1.75 & 84 & 59\\
1.87 & 2.34 & 3 & 2.24 & 162 & 94 & 6 & 2.19 & 316 & 129 & 2 & 2.29 & 114 & 81\\
2.34 & 2.93 & 1 & 2.46 & 51 & 51 & 5 & 2.62 & 293 & 131 & 0 & - & - & -\\
2.93 & 3.67 & 1 & 3.27 & 76 & 76 & 1 & 3.27 & 76 & 76 & 0 & - & - & -\\
3.67 & 4.58 & 3 & 3.92 & 249 & 144 & 5 & 3.92 & 414 & 185 & 0 & - & - & -\\
4.58 & 5.72 & 0 & - & - & - & 0 & - & - & - & 0 & - & - & -\\
5.72 & 7.15 & 3 & 6.30 & 412 & 238 & 3 & 6.30 & 412 & 238 & 0 & - & - & -\\
7.15 & 8.94 & 0 & - & - & - & 1 & 7.97 & 172 & 172 & 0 & - & - & -\\
8.94 & 11.14 & 0 & - & - & - & 0 & - & - & - & 0 & - & - & -\\
11.14 & 14.02 & 1 & 12.89 & 193 & 193 & 1 & 12.89 & 193 & 193 & 0 & - & - & -\\

\hline
\end{tabular}
}
\caption{Table of Euclidean-weighted differential source counts of different sub-populations of strongly scintillating population. Fig. \ref{Fig:compare_SubPopn_Counts} compares these counts against the GLEAM number counts of entire population of extragalactic sources at 162 MHz. The columns are as follows: $(1)$ Lower end of bin range (Jy); $(2)$ Higher end of the bins range (Jy); $(3)$ The raw counts for peaked spectrum sources; $(4)$ Median flux density of all sources in the bin (Jy) used to Euclidean-weight the counts; $(5)$ The differential number counts, normalised to the value expected for a Euclidean source counts distribution for peaked spectrum sources; $(6)$ The uncertainty in counts; $(7)$ The raw counts for peaked spectrum and CSS sources combined; $(8)$  Median flux density of all sources in the bin (Jy) used to Euclidean-weight the counts;  $(9)$ The differential number counts, normalised to the value expected for a Euclidean source counts distribution for peaked spectrum and CSS sources combined; $(10)$ The uncertainty in counts; $(11)$ The raw counts for flat-spectrum sources; $(12)$ Median flux density of all sources in the bin (Jy) used to Euclidean-weight the counts;  $(13)$ The differential number counts, normalised to the value expected for a Euclidean source counts distribution for flat-spectrum sources; $(14)$ The uncertainty in counts.}

\label{Tab:SrcCount_SubPopulations}
\end{table*}

\section{Summary}
\label{Sec:Conclusions}

We have developed a technique to derive source counts directly from the variability image developed in Paper I, while accounting for the changing sensitivity across the field of view.
Using this technique, we derive differential number counts of sub-arcsecond compact sources identified down to $S_{162\,MHz} = 0.4$\,Jy using widefield IPS with the MWA. This technique will be important to make efficient study of sub-arcsecond counts, with improved statistics, with the future IPS survey covering a large part of the sky (Morgan et al. in prep.). In the longer term, we envisage the technique extending counts of sub-arcsecond sources to microJy level using the SKA\_Low.

We show that the counts of sources with compact structures follows the behavior of the overall extragalactic population above $\sim$ 3\,Jy, and shows a Euclidean behavior over $\sim$ an order of magnitude in flux density below $\sim$3 \,Jy. We find an excellent agreement with the counts made using the S$^3$ sky model. We estimate the source density of compact objects, that can show some visibility to $\sim$350 km baselines at 0.4 Jy level to be 0.7 sources per square degree in the sky. Extending this to lower flux densities using the S$^{3}$ model, we estimate a surface density of 4.6 potential calibrator sources per square degrees at 0.04\,Jy usable for ionospheric calibration of the SKA\_Low. We show that the number of high-quality calibrators that can be identified using flat spectrum \emph{or} peaked spectrum as the compactness criterion approximately triples the number of good candidates relative to using flat spectrum alone.

Using the criteria of strong scintillation (NSI\,$\geq$\,0.90) we identify 93 point sources, unassociated with extended structures, and derive their counts. We, thus, probe an order of magnitude deeper in flux density than any previous counts of sub-arcsecond sources at low radio frequencies. The counts of point sources show an overall Euclidean behavior which is different from the counts derived for sources in the extragalactic GLEAM survey, thought to be dominated by radio galaxies. We show that these counts are a combination of counts of two sub-populations and that they will have to be investigated separately in order to understand their evolution. We find a strong difference between our counts of point sources and the counts predicted by the S$^3$ simulation.

We use an existing catalogue to identify peaked spectrum sources within our sample and divide the remainder of the point sources into flat-spectrum sources and compact steep-spectrum (CSS) sources. We note that the latter can only be identified via IPS or another high-resolution technique, since they are inseparable from the vast majority of low-frequency radio sources on the basis of the spectrum alone. We find that 81\% of the point source population is made up of the peaked-spectrum/CSS sources, when we combine them on the basis that both are thought to be the early stages in the evolution of radio galaxies. The remaining 19\% are flat-spectrum sources, with characteristics of synchrotron self-absorbed cores of radio galaxies. We present counts of these different sub-populations; however with our current data it is not possible to identify differences between the sub-populations with any confidence. We expect this to change once we have IPS data from a wider area of the sky.

\section*{Acknowledgements}

This scientific work makes use of the Murchison Radio-astronomy Observatory, operated by CSIRO. We acknowledge the Wajarri Yamatji people as the traditional owners of the Observatory site. Support for the operation of the MWA is provided by the Australian Government (NCRIS), under a contract to Curtin University administered by Astronomy Australia Limited. We acknowledge the Pawsey Supercomputing Centre which is supported by the Western Australian and Australian Governments.
Parts of this research were conducted by the Australian Research Council Centre of Excellence for All-sky Astrophysics (CAASTRO), through project number CE110001020. 
We used the TOPCAT software \citep[]{Taylor2005} for the analysis of some of our data.



\bibliographystyle{mnras}
\bibliography{references}

\end{document}